# Self-organization of weighted networks for optimal synchronizability

Louis Kempton,[1,*] Guido Herrmann,[2,†] and Mario di Bernardo[3,4,‡]

[1] *Bristol Centre for Complexity Sciences, University of Bristol, UK*
[2] *Department of Mechanical Engineering, University of Bristol, UK*
[3] *Department of Engineering Mathematics, University of Bristol, UK*
[4] *Department of Electrical Engineering and Information Technology, University of Naples Federico II, Italy*

We show that a network can self-organize its structure in a completely distributed manner in order to optimize its synchronizability whilst satisfying the local constraints: non-negativity of edge weights, and maximum weighted degree of nodes. A novel multilayer approach is presented which uses a distributed strategy to estimate two spectral functions of the graph Laplacian, the algebraic connectivity $\lambda_2$ and the eigenratio $r = \lambda_n/\lambda_2$. These local estimates are then used to evolve the edge weights so as to maximize $\lambda_2$, or minimize $r$ and, hence, achieve an optimal structure.

Many networked systems that we see in nature adapt their structure in time to better perform a specific function. Learning through changing the strength of synapses [1], or through rewiring of a collection of neurons [2] can be seen as a paragon of this observation. For such networks, the global behaviour emerges from the local interactions of agents and it is these agents which can adapt according to their local environment to steer the macroscopic network behaviour and functionality. Hence, devising microscopic strategies to steer the macroscopic behaviour of physical networks in a controlled way can offer a number of benefits in engineering applications from the synchronization of power networks [3] to the coordination of robotic swarms [4]. Adaptive networks can be robust, coping well with missing or broken parts; they are able to self-organize, removing the requirement for a dedicated designer, and adapting to changes in the operating environment in real time; moreover these systems tend to scale well with an increasing number of agents [5].

It is the behaviour of self-organization which motivates this Letter. In particular, we ask, "Can networks adapt into globally optimal structures, using only deterministic local feedback interactions between agents?". In our model, we focus on the problem of finding the weighted network with globally optimal synchronizability, defined alternatively as maximizing the algebraic connectivity $\lambda_2$, the second smallest eigenvalue of the graph Laplacian, or by minimizing the eigenratio $\lambda_n/\lambda_2$, with $\lambda_n$ being the largest Laplacian eigenvalue. We add some constraints to model the realistic case of each agent having a limited communication bandwidth, with edges only being physical with positive weight. Specifically, we will upper bound the weighted degree of each node by a constant value, and require that edge weights be non-negative. A real world example of these constraints are the limitations of the shared link bandwidths common in wired and wireless sensor and actuator networks (e.g., in disitributed robot swarms), where distributed optimization of the network can be of great benefit; see, for example, [6].

These two problems, maximizing $\lambda_2$ and minimizing $\lambda_n/\lambda_2$, have been well studied previously, with many techniques proposed, both for weighted and unweighted, directed and undirected networks. However, previous methods either require global knowledge of the network to find the optimal structure, or optimality is not guaranteed when local rules are used solely. In the first case, finding the structure which maximizes algebraic connectivity can be formulated as a semi-definite program (SDP) [7] and extended to the unweighted case using mixed integer semi-definite programming (MISDP) [8, 9]; in this case, it is assumed that the solver has complete knowledge of the network. Likewise, methods employing simulated annealing with edge-rewiring, to find optimal, or near-optimal, graphs [10–12] also require an external supervisor of the network with global knowledge to evolve the network structure. In the second case, only local information can be used to assign edge weights: for example, node degree may be used to modify edge weights as in [13], enhancing $\lambda_n/\lambda_2$, but convergence towards an optimal structure cannot be guaranteed.

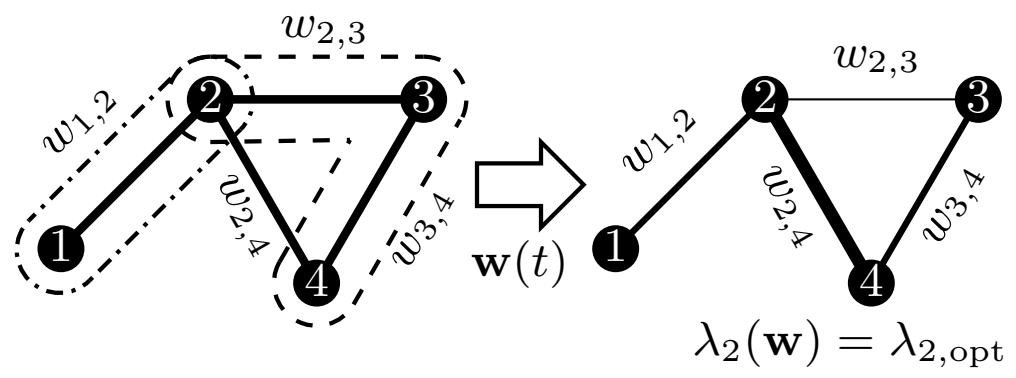

Figure 1. A schematic diagram of the problem where we wish to adapt edge weights $\mathbf{w} = [w_{i,j}]_{m\times 1}$ in time to maximize $\lambda_2$ of a given graph with $n$ nodes and $m$ edges, in a distributed manner. For clarity, the horizon of knowledge for edge $\{1, 2\}$ (dot-dash) and node 3 (dashed) are shown. Different values of the edge weights are represented in terms of thickness of the network links.

In this Letter, we present a new optimal strategy to evolve the edge weights so as to maximize $\lambda_2$ or minimize $\lambda_n/\lambda_2$ for a given undirected graph. Unlike previous methods which find the optimal network structure, we impose the further constraint of only using information local to each node to adapt the network, see Figure 1 for a schematic of the problem and the horizons of each agent's knowledge. Namely, nodes may only com-municate with their one-hop neighbours. The resulting strategy employs a multi-layer framework, and drives the network to the globally optimal structure in a fully decentralized manner.

We consider the generic homogeneous networked system of $n$ agents:

$$\dot{\mathbf{x}}_i = \mathbf{F}(\mathbf{x}_i) - \sigma \sum_{j=1}^{n} l_{i,j} \mathbf{H}(\mathbf{x}_j) \tag{1}$$

where $\mathbf{x}_i = [x_j]_{d\times 1}$ is the $d$-dimensional state vector of the $i^{\text{th}}$ system, $\mathbf{F}(\mathbf{x}_i) : \mathbb{R}^d \mapsto \mathbb{R}^d$ is the vector field of the $i^{\text{th}}$ isolated system, $\sigma$ is a scalar global coupling parameter, $\mathbf{H}(\mathbf{x}_j) : \mathbb{R}^d \mapsto \mathbb{R}^d$ is the coupling function between nodes, and $\mathbf{L} = [l_{i,j}]_{n\times n}$ is the weighted graph Laplacian, defined as a function of edge weights $w_{i,j}$ as:

$$l_{i,j} = \begin{cases} \sum_{j\in\mathcal{N}_i} w_{i,j} & \text{if } i = j \\ -w_{i,j} & \text{if } i \neq j \end{cases} \tag{2}$$

where $\mathcal{N}_i$ is the one-hop neighbourhood of node $i$, so that $\mathbf{L}$ is a square symmetric matrix with zero row sum. The eigenvalues of the graph Laplacian are thus real, and can then be ordered $0 = \lambda_1 < \lambda_2 \leq \cdots \leq \lambda_n$.

For populations of such identical, coupled, nonlinear systems, the predominant technique for determining the local stability of the synchronous solution is the Master Stability Function (MSF) approach [14, 15]. In general, for a given MSF $\Psi(\alpha)$, intervals in the scalar argument $\alpha$ for which its value is negative fall into two categories: right-unbounded $\Psi(\alpha) < 0$ for $\alpha \in (\alpha_1, \infty)$, or proper bounded $\Psi(\alpha) < 0$ for $\alpha \in (\alpha_1, \alpha_2)$. Using the classification system of Huang et al. [15], those MSFs which contain only one negative interval and that interval is right-unbounded are called $\Gamma_1$, and those for which their sole negative interval is proper bounded are called $\Gamma_2$. For systems with a $\Gamma_1$ MSF, the synchronous solution is stable if $\sigma\lambda_2(\mathbf{L}) > \alpha_1$ (Case 1). That is, networks with greater algebraic connectivity $\lambda_2$ require a lower global coupling strength, and are thus more easily synchronized. On the other hand, for systems with a $\Gamma_2$ MSF, the synchronous solution can only be stable if $\alpha_1 < \sigma\lambda_2 \leq \sigma\lambda_n < \alpha_2$ (Case 2). Thus, such systems can only permit a locally transversally stable synchronous solution if $\lambda_n/\lambda_2 < \alpha_2/\alpha_1$. Graphs with lower eigenratio $r \coloneqq \lambda_n/\lambda_2$ are then deemed more synchronizable [16]. Thus, depending on the shape of the specific MSF for the networked system, one of these two spectral functions of the graph Laplacian determines its synchronizability.

If we wish to maximize synchronizability, then we must consider two cases, which can be formulated in the standard form of the optimization problem

$$\begin{aligned} &\underset{\mathbf{w}}{\text{minimize}} && f(\mathbf{w}) \\ &\text{subject to} && \mathbf{w} \in \mathcal{W} \end{aligned} \tag{3}$$

where $\mathbf{w} = [w_{i,j}]_{m\times 1}$ is the vector of all edge weights, the objective function $f(\mathbf{w}) = -\lambda_2$ for Case 1 or $f(\mathbf{w}) = r$ for Case 2, and $\mathcal{W}$ is a set of feasible edge weights, which is both closed and convex. Here, we will use the following feasible region,

$$\mathcal{W} = \{\mathbf{w} : \mathbf{w} \geq \mathbf{0} \wedge l_{i,i} \leq k_i, \forall i\} \tag{4}$$

so as to allow only non-negative edge weights, and upper bound the weighted degree of each node by a constant $k_i$ (feasible edge weights lie in a closed polytope in the positive orthant of $\mathbb{R}^m$).

To optimize the desired objective (maximizing $\lambda_2$ or minimizing $r$) we take advantage of the fact that $\lambda_2(\mathbf{L})$ is a concave function of the edge weights [7] ($-\lambda_2(\mathbf{L})$ is then a convex function), and $r(\mathbf{L})$ is a quasiconvex function of edge weights [17], thus for either function any local minimizer is a global minimizer. Then as our set of feasible edge weights is a convex set, we can minimize $-\lambda_2(\mathbf{L})$ or $r(\mathbf{L})$ by gradient descent in a distributed fashion. This gradient descent method forms the top level in a hierarchy of distributed processes (see Figure 2) which is added to an underlying layer designed to estimate the gradient of the chosen objective. Note that Case 2 is an example of a single ratio quasi-convex fractional program, and thus can be transformed to a parameter-free convex program, as described in [17] Proposition 8.

First, let us describe the *Weight optimizer layer* in Figure 2. The goal of this layer is to minimize the objective function $f(\mathbf{w})$ by forcing edge weights in the direction of steepest descent of a modified objective function $g(\mathbf{w})$, which enforces the boundary constraints of the feasible region through the use of logarithmic barriers [18]:

$$g(\mathbf{w}) = f(\mathbf{w}) - \frac{1}{q(t)} \left( \sum_{\{i,j\}\in\mathcal{E}} \log(w_{i,j}) + \sum_{i=1}^{n} \log(k_i - l_{i,i}) \right) \tag{5}$$

For conciseness of notation we have used $\mathcal{E}$ to signify the edge set, so that the first summation is over all edges in the network. The strength of these barriers is determined by the function $q(t)$ which is chosen to be positive monotonic increasing and unbounded. It increases as soon as the minimal value of $f(\mathbf{w})$ is approached (see the supplementary material for the choice of $q(t)$). This guarantees that the bounds of $\mathcal{W}$ are kept, while converging to the (constrained) optimum $\mathbf{w}_{\text{opt}}$.

Each edge weight is then adapted in time according to:

$$\ddot{w}_{i,j} = -k_a \frac{\partial g(\mathbf{w})}{\partial w_{i,j}} - c_1 \dot{w}_{i,j} \tag{6}$$

where the sensitivity of the modified objective with respect to an edge weight can be computed as:

$$\frac{\partial g(\mathbf{w})}{\partial w_{i,j}} = \frac{\partial f(\mathbf{w})}{\partial w_{i,j}} - \frac{1}{q(t)} \left( \frac{1}{w_{i,j}} - \frac{1}{k_i - l_{i,i}} - \frac{1}{k_j - l_{j,j}} \right) \tag{7}$$

It can clearly be seen that the forcing from the logarithmic barrier functions on the adaptive dynamics of a single edge requires only information local to that edge (its own weight $w_{i,j}$, the weighted degrees of its parents $l_{i,i}$ and $l_{j,j}$ and their maximum allowed weighted degrees $k_i$ and $k_j$). Only the sensitivity of the chosen objective function, $\frac{\partial f(\mathbf{w})}{\partial w_{i,j}}$, with respect to the edge remains as a global parameter, which will be computed locally via a set of distributed estimators (in Case 1, by the $\lambda_2$ Estimator layer, shown in Figure 2). This will permit the optimization algorithm to be fully distributed.

In particular, to estimate the sensitivities of the algebraic connectivity to variation of the edge weights, we use the distributed strategy by Yang et al. [19] to evaluate the algebraic connectivity of a weighted undirected network in a distributed fashion. The strategy can be implemented as two additional layers, the *Proportional-Integral (PI) Consensus layer* and the $\lambda_2$*-Estimator layer* shown in Fig. 2. The dynamics of these two layers can be described by the following set of differential equations inspired by power iteration (see [19] for further details) :

$$\dot{\mathbf{a}} = -k_1\boldsymbol{\varphi}_{\mathbf{a}} - k_2\mathbf{L}\mathbf{a} - k_3(\boldsymbol{\psi}_{\mathbf{a}} - \mathbf{1}) \circ \mathbf{a} \tag{8}$$

$$\dot{\boldsymbol{\varphi}}_{\mathbf{a}} = \gamma(\mathbf{a} - \boldsymbol{\varphi}_{\mathbf{a}}) - k_P\mathbf{L}\boldsymbol{\varphi}_{\mathbf{a}} - k_I\mathbf{L}\boldsymbol{\chi}_{\mathbf{a}} \tag{9}$$
$$\dot{\boldsymbol{\chi}}_{\mathbf{a}} = k_I\mathbf{L}\boldsymbol{\varphi}_{\mathbf{a}}$$

$$\dot{\boldsymbol{\psi}}_{\mathbf{a}} = \gamma(\mathbf{a}^2 - \boldsymbol{\psi}_{\mathbf{a}}) - k_P\mathbf{L}\boldsymbol{\psi}_{\mathbf{a}} - k_I\mathbf{L}\boldsymbol{\omega}_{\mathbf{a}} \tag{10}$$
$$\dot{\boldsymbol{\omega}}_{\mathbf{a}} = k_I\mathbf{L}\boldsymbol{\psi}_{\mathbf{a}}$$

For concise notation, component-wise product of vectors is signified by $\circ$, and squaring a vector is taken component-wise also, so that $\mathbf{a}^2 = \mathbf{a} \circ \mathbf{a}$. Here $\mathbf{a}$ is an estimate of the eigenvector associated with $\lambda_2$, which requires two further global variables: the arithmetic mean of $\mathbf{a}$, $\langle\mathbf{a}\rangle \triangleq 1/n \sum a_i$, and the mean of the squared components, $\langle\mathbf{a}^2\rangle \triangleq 1/n \sum a_i^2$.

These global variables are estimated in a distributed manner using a further layer (see Figure 2) consisting of two Proportional-Integral (PI) consensus estimators [20], with $\boldsymbol{\varphi}_{\mathbf{a}}$ being an estimate of $\langle\mathbf{a}\rangle\mathbf{1}$, and $\boldsymbol{\psi}_{\mathbf{a}}$ being an estimate of $\langle\mathbf{a}^2\rangle\mathbf{1}$.

The parameters $k_1$, $k_2$ and $k_3$ control three actions which can be summarized as deflation, direction update, and renormalization, respectively [19]. The result of these actions is that for $\mathbf{a}$, there are two stable stationary points, which together are global attractors if $k_1 > k_3 \geq k_2\lambda_n$:

$$\mathbf{a}^* = \pm\mathbf{v_2}(\mathbf{L})\sqrt{\frac{n(k_3 - k_2\lambda_2)}{k_3}} \tag{11}$$

where $\mathbf{v_2}$ is the unit eigenvector associated with $\lambda_2$. Hence, $\lambda_2$ can be estimated by node $i$ using Equations (8) to (11) as:

$$\widehat{\lambda_2}^{(i)} = \frac{k_3}{k_2}(1 - \psi_{\boldsymbol{a}i}) \tag{12}$$

By reversing the sign of the direction update in (8), this distributed $\lambda_2$ estimator may be modified to form a distributed estimator for $\lambda_n$ and the estimated eigenvector $\mathbf{b}$ associated with it (see Equation (3), supplementary material, for details).

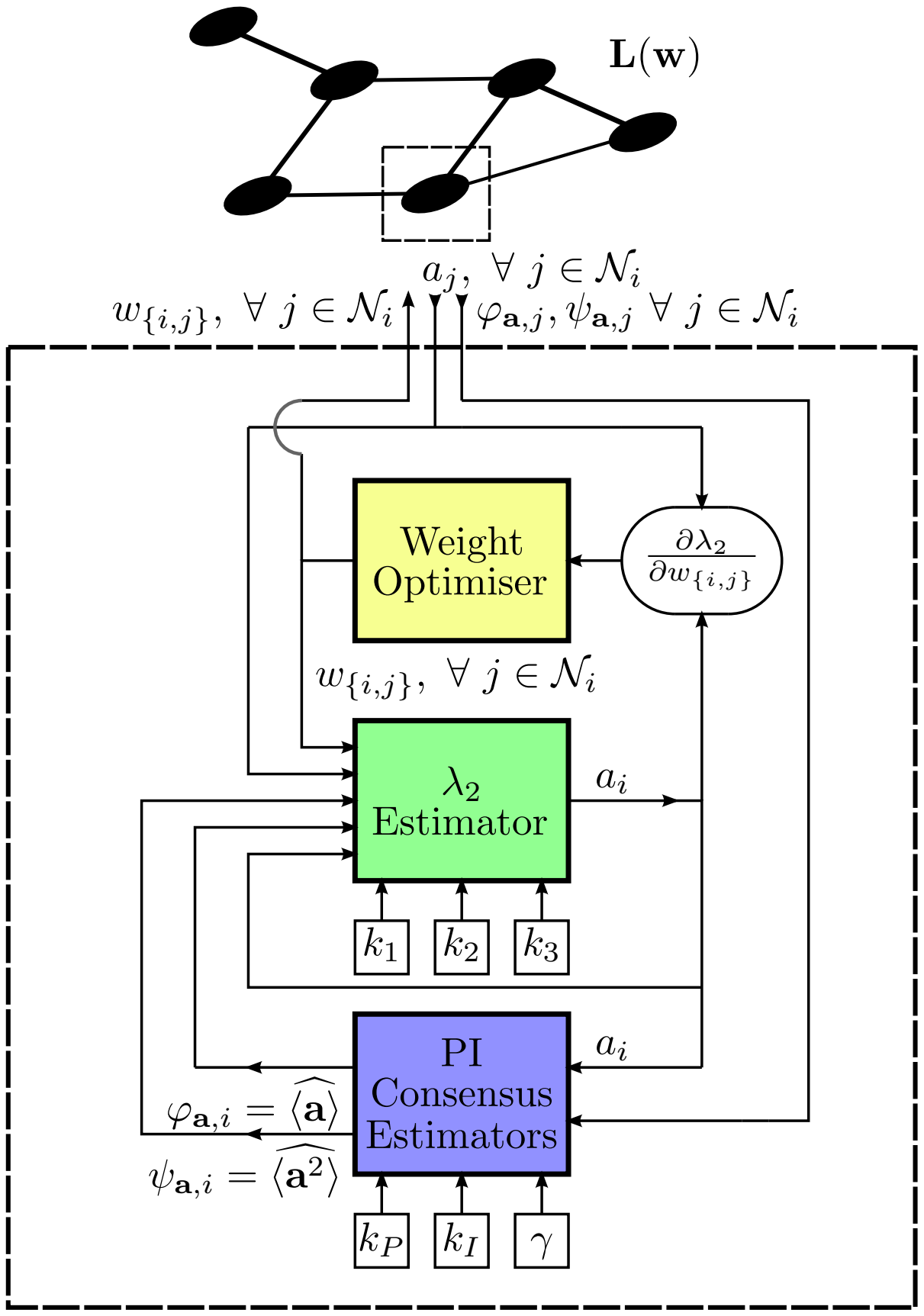

Figure 2. Schematic diagram of the distributed multilayer approach for $\lambda_2$ maximisation proposed in the paper, with faster processes at the bottom, and slower processes built on top. The processes occurring in a single node are expanded, and all nodes follow identical rules. For the corresponding diagram for $r$ minimisation, see the supplementary material.

These methods to make distributed estimates of $\lambda_2$ and $\lambda_n$ can further be used by each node to obtain local estimates of the sensitivity of each function with respect to an edge weight $w_{i,j}$ (derivations can be found in the supplementary material):

$$\widehat{\frac{\partial\lambda_2}{\partial w_{i,j}}}^{(i)} = \frac{(a_i - a_j)^2}{n\psi_{\boldsymbol{a}i}} \tag{13}$$

$$\widehat{\frac{\partial\lambda_n}{\partial w_{i,j}}}^{(i)} = \frac{(b_i - b_j)^2}{n\psi_{\boldsymbol{b}i}} \tag{14}$$

Moreover, it is simply a matter of applying the quotient rule for differentiation to estimate the sensitivity of $r$:

$$\widehat{\frac{\partial r}{\partial w_{i,j}}}^{(i)} = \frac{\widehat{\lambda_2}^{(i)}\widehat{\frac{\partial\lambda_n}{\partial w_{i,j}}}^{(i)} - \widehat{\lambda_n}^{(i)}\widehat{\frac{\partial\lambda_2}{\partial w_{i,j}}}^{(i)}}{\left(\widehat{\lambda_n}^{(i)}\right)^2} \tag{15}$$

It is now possible to estimate the gradient of either objective function in a distributed manner, and hence the edge weights can be adapted by steepest descent, whilst satisfying the feasibility constraints, also in a distributed fashion. It is worth noting that we require the estimators to converge faster than the weights adapt so that the sensitivities can be estimated with enough accuracy.

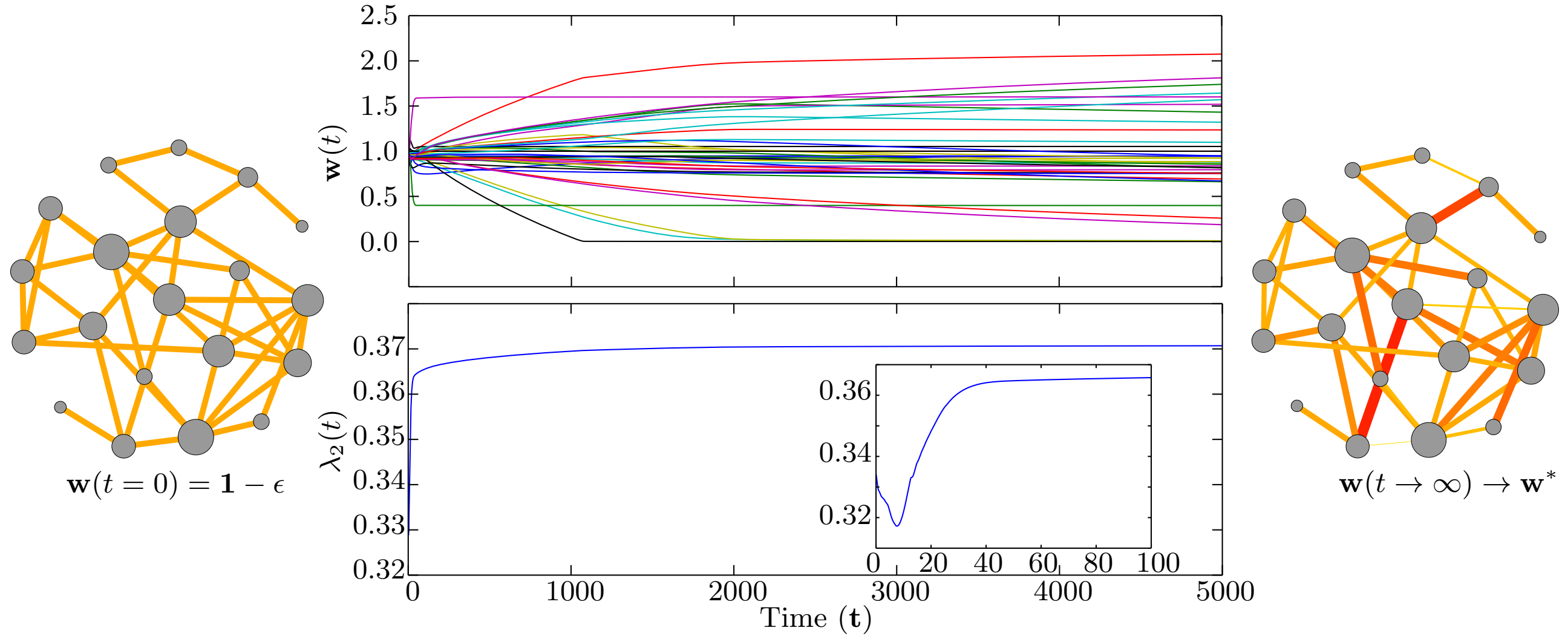

Figure 3. Edge weights are adapted in time according to our distributed algorithm. The algebraic connectivity $\lambda_2$ of the network increases over time and settles to a maximum value. To see the initial dip in connectivity while the layers of estimators take time to converge, see the inset. In the diagrams showing the initial and end state of the network, node diameter is proportional to maximum allowed weighted degree, $k_i$, and edge thickness and colour is proportional to weight, higher weights are redder and thicker. Case 2, when $f(\mathbf{w}) = r$, can be found in the supplementary material.

In Figure 3, we illustrate a representative application of our strategy to adapt the edges of a small undirected network for maximizing the algebraic connectivity $\lambda_2$, Case 1. We have randomly generated a connected graph of twenty nodes, and set all weights at $w_{i,j}(0) = 1 - \epsilon$, where $\epsilon$ is small so that the initial edge weights lie in the interior of the feasible set. When all weights are equal to 1, the initial algebraic connectivity is found to be $\lambda_2(0) \approx 0.3344$. Edge weights are then adapted according to the algorithm in Figure 2, using Equations (8) to (10) with (13) to estimate the gradient of the objective function, and (6) to adapt each edge weight according to gradient descent. We choose control parameters so that the PI consensus estimator layer is approximately 10 times faster than the algebraic connectivity estimator, which in turn is chosen to be 10 times faster than the weight adaptation layer. This gives us sufficient separation in time scale between the layers. To bound the feasible set of edge weights, no weight is allowed to be negative, and the maximum allowed weighted degree of each node $k_i$ is chosen to be the degree of the node. Through this choice of a feasible set, we can be sure that total weight in the network may not increase over time, and any improvement in $\lambda_2$ must be due to better distribution of edge weight, rather than absolute increase of their total value.

As edge weights are adapted, the algebraic connectivity initially decreases until the estimator layers have properly converged, whereupon it increases rapidly, as the nadir of the potential well of the objective function for a given $q$ is reached. Finally, as $q$ increases and the logarithmic barriers enforcing the feasible set become steeper, the edge weights slowly converge to their optimal values.

It should be noticed that the algebraic connectivity converges more rapidly than the slowest converging edges due to weak sensitivity of $\lambda_2$ with respect to some edges. After 5000 seconds of simulated time, the algebraic connectivity has reached a value of $\lambda_2(5000) \approx 0.3707$, which over 99.9% of the result for optimal $\lambda_2$ found using the method of [7]: $\lambda_2^* \approx 0.3708$. At this time, some of the edges have yet to converge, but it is known that three edges will tend to zero value, and could be removed from the network at no detriment to the algebraic connectivity.

In Figure S4 of the supplementary material, we show the edge adaptation for the alternative optimization problem described in Case 2, minimization of the eigenratio $r$. For this particular case we highlight that edges do not converge to a final value if the optimal network has non-distinct eigenvalues in the objective function $f(\mathbf{w})$. Indeed, in the optimal network for the given graph and constraints, $\lambda_n \approx \lambda_{n-1} \approx \lambda_{n-2}$, see Figure S5 in the supplementary material for the trajectories of the eigenvalues of the weighted graph Laplacian.

In this Letter, we have shown that a network of agents may cooperate together, exchanging only local knowledge, and come to agreement on a globally optimal network structure, through the use of a multi-layer weight adaptation algorithm. By means of such a multilayer approach, we showed that it is possible for nodes in the network to locally estimate the sensitivities of two global spectral functions of the graph Laplacian: the algebraic connectivity $\lambda_2$ and the eigenratio $\lambda_n/\lambda_2$. This information can then be used by the edges to locally adapt their weights so as to maximize the network synchronizability.

It is therefore possible for the network to self-organize in order to steer some macroscopic observables, such the algebraic connectivity, by using only microscopic information. Future work will be aimed at investigating how to extend this approach to control other emerging properties of a network of interest and, hence, achieve global control of the network via local adaptive rules.

---

* l.kempton@bristol.ac.uk
† g.herrmann@bris.ac.uk
‡ m.dibernardo@bris.ac.uk

# Supplementary Material for "Self-organization of weighted networks for optimal synchronizability"

Louis Kempton,[1,*] Guido Herrmann,[1,†] and Mario di Bernardo[1,2,‡]
[1] *University of Bristol, UK*
[2] *University of Naples Federico II, Italy*

## THE RÖSSLER OSCILLATOR & MSF CHARACTERISTIC SHAPES

As a representative example, we consider a network of coupled Rössler oscillators with parameters as in [1]. In the fashion of Equation (1) in the main document, the isolated system is given by

$$\mathbf{x} \triangleq [x, y, z]^\top$$
$$\dot{\mathbf{x}} = \mathbf{F}(\mathbf{x}) \triangleq \begin{bmatrix} -y - z \\ x + \zeta y \\ \beta + z(x - \gamma) \end{bmatrix}, \qquad \zeta = 0.2, \beta = 0.2, \gamma = 9$$

and we investigate two different coupling regimes:

$x$-diffusive coupling:

$$\mathbf{H}(\mathbf{x}) \triangleq [x, 0, 0]^\top$$

$y$-diffusive coupling:

$$\mathbf{H}(\mathbf{x}) \triangleq [0, y, 0]^\top$$

These two coupling schemes yield Master Stability Functions of class $\Gamma_2$ and $\Gamma_1$ respectively:

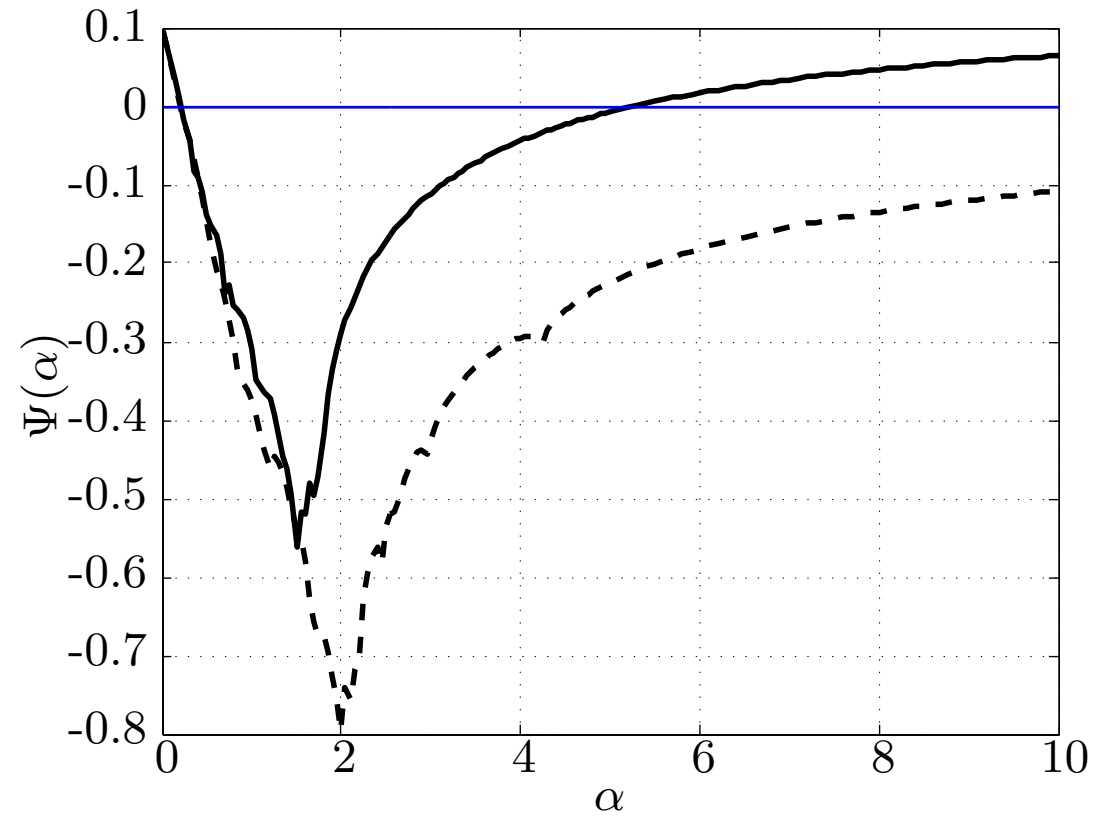

FIG. S 1: The MSFs $\Psi(\alpha)$ for the Rössler oscillator with two different couplings are shown, showing class $\Gamma_1$ ($y$-diffusive coupling, dashed line) and $\Gamma_2$ ($x$-diffusive coupling, solid line) characteristic shapes.

The two characteristic shapes of a negative interval in an MSF can be seen. Under $x$-diffusive coupling the corresponding MSF (solid line) crosses the x axis twice, creating a proper-bounded negative interval. In a network of such $x$-diffusively coupled Rössler oscillators, all eigenvalues of the graph Laplacian scaled by the global coupling constant must fall into this negative interval for local stability of the synchronous manifold. This class of MSF leads to the idea of minimizing the ratio $\lambda_n/\lambda_2$ to improve synchronizability [1, 2]. On the other hand, under $y$-diffusive coupling the corresponding MSF (dashed line) is right unbounded. For a connected network of Rössler oscillators coupled in this manner, the synchronous manifold can always be stabilised with sufficiently high global coupling parameter. However, the larger the algebraic connectivity $\lambda_2$ is, the lower the global coupling parameter may be, whilst still guaranteeing local stability of the synchronous solution. Thus, the network synchronizability can be improved by increasing $\lambda_2$.

## CHOICE OF STEEPNESS FUNCTION $q(t)$

The basic idea of the weight adaption layer, Equation (6) in the main text, is that edge weights will evolve to the minimum of a potential well $g(\mathbf{w})$. If we consider a constant function $q(t)$ in $g(\mathbf{w})$, this is evident as Equation (6) is a dissipative dynamical system.

However, matters are slightly complicated by the notion that the logarithmic barrier functions of the well are required to become steeper in time, through the action of $q(t)$. It is sufficient that $q(t)$ be positive, monotonic increasing, unbounded function, that does not escape to infinity in finite time, for edge weights to converge to the minimum of the desired objective $f(\mathbf{w})$. However, we choose to increase the steepness of the barrier functions as the current minimum of the well is approached. Conceptually, the steepness of the logarithmic barriers could be increased in proportion to the current flatness of the well, using the adaptive law:

$$\ddot{q} = \frac{k_b}{||\nabla g(\mathbf{w})||_2 + \delta} - c_2\dot{q} \tag{1}$$

Unfortunately this equation is not fully distributed, violating our requirements. To overcome this problem we assign assign a local $q_{i,j}$ for each weight, and update according to:

$$\ddot{q}_{i,j} = \frac{k_b}{||\frac{\partial g(\mathbf{w})}{\partial w_{i,j}}||_2 + \delta} - c_2\dot{q}_{i,j} \tag{2}$$

The positive control parameters $k_b$ and $c_2$ can be tuned to determine the aggressiveness of the barrier steepening, while the control parameter $\delta$ (also positive) is used to avoid the singularity as edge weights approach their optimal values.

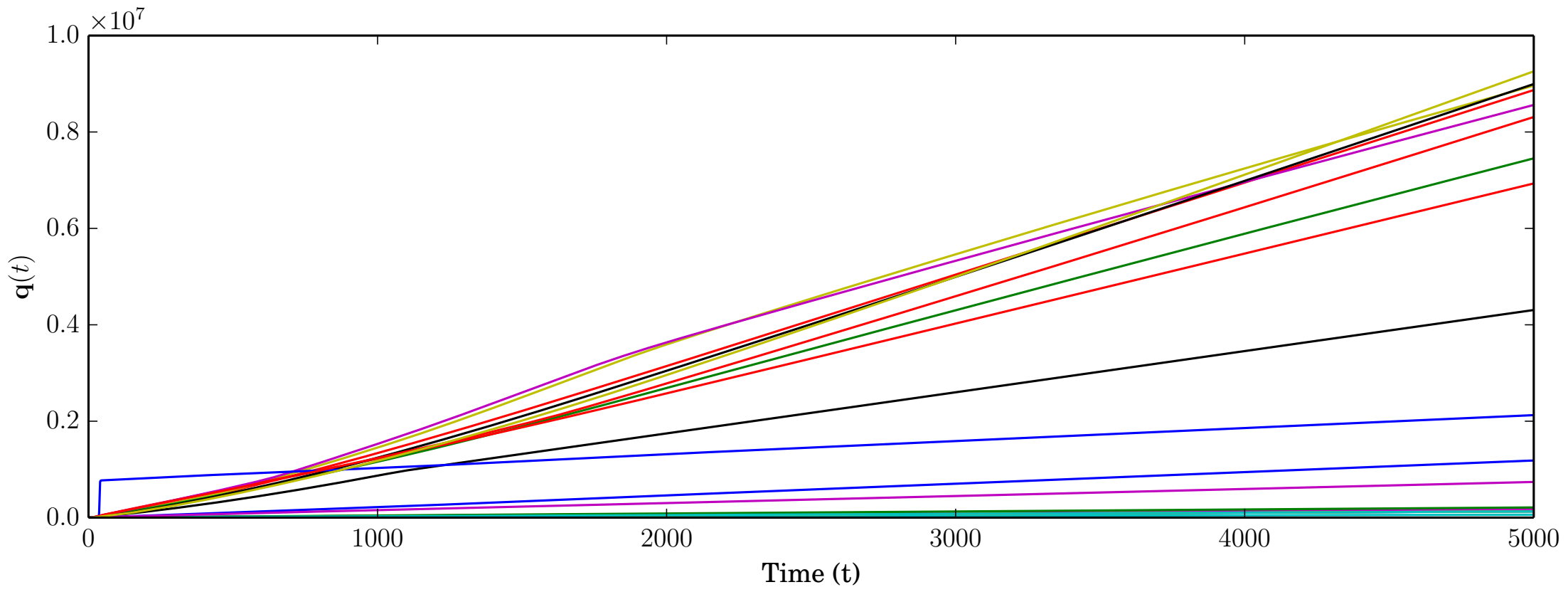

FIG. S 2: Here we show how the values of $q_{i,j}$ grow over time in the simulation described in the main text, maximizing the algebraic connectivity $\lambda_2$ of the network.

## DISTRIBUTED $\lambda_n$ ESTIMATION

The largest Laplacian eigenvalue can be estimated using a similar approach to the method from [3] presented in the main text to estimate $\lambda_2$. Specifically, we can use the equations:

$$\begin{aligned}
\dot{\mathbf{b}} &= -k_1\boldsymbol{\varphi_b} + k_2\mathbf{L}\mathbf{b} - nk_3(\boldsymbol{\psi_a} - \mathbf{1})\cdot\mathbf{b} \\
\dot{\boldsymbol{\varphi}}_{\mathbf{b}} &= \gamma(\mathbf{b} - \boldsymbol{\varphi_b}) - k_P\mathbf{L}\boldsymbol{\varphi_b} - k_I\mathbf{L}\boldsymbol{\chi_b} \\
\dot{\boldsymbol{\chi}}_{\mathbf{b}} &= k_I\mathbf{L}\boldsymbol{\varphi_b} \\
\dot{\boldsymbol{\psi}}_{\mathbf{b}} &= \gamma(\mathbf{b}^2 - \boldsymbol{\psi_b}) - k_P\mathbf{L}\boldsymbol{\psi_b} - k_I\mathbf{L}\boldsymbol{\omega_b} \\
\dot{\boldsymbol{\omega}}_{\mathbf{b}} &= k_I\mathbf{L}\boldsymbol{\psi_b}
\end{aligned} \tag{3}$$

so that $\mathbf{b}$ will converge to an eigenvector associated with $\lambda_n$. Node $i$ can then compute a local estimate of $\lambda_n$:

$$\widehat{\lambda_n}^{(i)} = \frac{nk_3}{k_2}(\psi_{\boldsymbol{b}i} - 1) \tag{4}$$

This process can be seen as the mirror to the algebraic connectivity estimator presented in [3]. Instead of the term associated with $k_2$ contracting the state towards consensus so that the slowest mode dominates, it now expands away from consensus so that the fastest mode dominates. Again, $k_1$ ensures that the estimate is perpendicular to the consensus mode and $k_3$ acts to renormalize and stop the estimate from diverging.

## SENSITIVITY DERIVATION

Here we derive, in a similar manner to [4], the partial derivatives of a generic eigenvalue, say $\lambda$, of the weighted graph Laplacian with respect to the edge weight $w_{i,j}$. Imagine that we know an associated right eigenvector $\mathbf{v}$ so that we can define the unit eigenvector $\hat{\mathbf{v}} = \mathbf{v}/||\mathbf{v}||_2$. As $\mathbf{L}$ is symmetric, we know that the unit left eigenvector is the transpose $\hat{\mathbf{u}} = \hat{\mathbf{v}}^\top$. Then, pre-multiplying the eigenvector relation $\mathbf{L}\hat{\mathbf{v}} = \lambda\hat{\mathbf{v}}$ by the unit left eigenvector yields:

$$\hat{\mathbf{v}}^\top \mathbf{L}\hat{\mathbf{v}} = \lambda\hat{\mathbf{v}}^\top\hat{\mathbf{v}} = \lambda \tag{5}$$

Taking the componentwise derivative with respect to the edge weight $w_{i,j}$, we find that:

$$\frac{\partial\lambda}{\partial w_{i,j}} = \frac{\partial\hat{\mathbf{v}}^\top}{\partial w_{i,j}}\mathbf{L}\hat{\mathbf{v}} + \hat{\mathbf{v}}^\top\frac{\partial\mathbf{L}}{\partial w_{i,j}}\hat{\mathbf{v}} + \hat{\mathbf{v}}^\top\mathbf{L}\frac{\partial\hat{\mathbf{v}}}{\partial w_{i,j}}$$

As $\mathbf{L}$ is symmetric, it is ensured that

$$\frac{\partial\hat{\mathbf{v}}^\top}{\partial w_{i,j}}\mathbf{L}\hat{\mathbf{v}} + \hat{\mathbf{v}}^\top\mathbf{L}\frac{\partial\hat{\mathbf{v}}}{\partial w_{i,j}} = \lambda\frac{\partial(\hat{\mathbf{v}}^\top\hat{\mathbf{v}})}{\partial w_{i,j}} = 0$$

Thus,

$$\frac{\partial\lambda}{\partial w_{i,j}} = \hat{\mathbf{v}}^\top\frac{\partial\mathbf{L}}{\partial w_{i,j}}\hat{\mathbf{v}} \tag{6}$$

Without loss of generality we can relabel nodes $i$ and $j$ to 1 and 2, revealing

$$\begin{aligned}\frac{\partial\lambda}{\partial w_{1,2}} &= \frac{1}{||\mathbf{v}||_2^2}\mathbf{v}^\top\left(\begin{array}{cc|c} +1 & -1 & \\ -1 & +1 & 0 \\ \hline \multicolumn{2}{c|}{0} & 0\end{array}\right)\mathbf{v} \\ &= \frac{1}{||\mathbf{v}||_2^2}\left(v_1 - v_2\right)^2 \end{aligned}\tag{7}$$

where $\mathbf{v} = [v_1, v_2, \ldots, v_n]^\top$.

From the distributed estimation procedure outlined in Equations (8) to (10) in the main text, and the set of Equations (3) in the supplementary text, we can compute estimates for the eigenvectors associated with $\lambda_2$ and $\lambda_n$, respectively $\mathbf{a}$ and $\mathbf{b}$, and estimates of the mean of their squared components, respectively $\boldsymbol{\psi_a}$ and $\boldsymbol{\psi_b}$. Using the relabelling argument that any two nodes could be labelled 1 and 2, along with Equation (7), we arrive at node $i$'s distributed estimates for the sensitivities:

$$\begin{aligned}\widehat{\frac{\partial\lambda_2}{\partial w_{i,j}}}^{(i)} &= \frac{(a_i - a_j)^2}{n\psi_{\mathbf{a}i}} \\ \widehat{\frac{\partial\lambda_n}{\partial w_{i,j}}}^{(i)} &= \frac{(b_i - b_j)^2}{n\psi_{\mathbf{b}i}}\end{aligned}$$

A schematic diagram of the distributed multi-layer approach for $r$ minimization is shown in Figure S3.

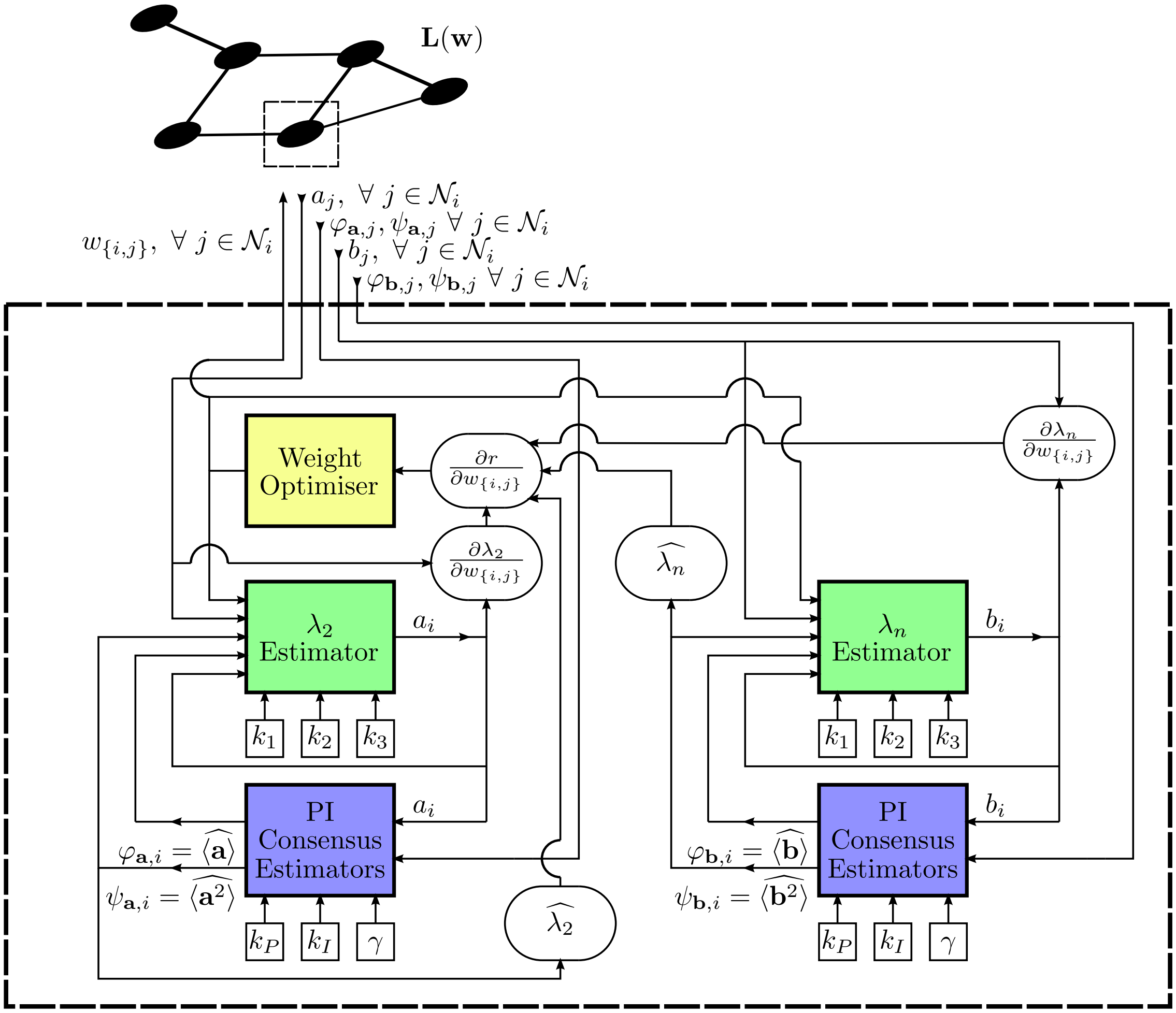

FIG. S 3: Schematic diagram of our distributed method for $r$ minimization. The flow diagram structure is a schematic of how variables interact in the different estimation layers within a single node $i$. On the left of the diagram, the algebraic connectivity is being estimated (green block) which requires two PI average consensus estimators (blue block). From these blocks, distributed estimates are made for $\lambda_2$ and $\partial\lambda_2/\partial w_{i,j}$ for each edge that connects to node $i$. In a similar fashion, on the right hand side of the diagram, distributed estimates of $\lambda_n$ and $\partial\lambda_n/\partial w_{i,j}$ are being made. Combining these estimates, a local estimate of $\partial r/\partial w_{i,j}$ is formed, and this is fed into the weight optimizer (yellow block), which uses this estimate and the local boundary constraints to inform the adaptation of each of the weights.

## EIGENRATIO OPTIMIZATION SIMULATION

We use the same example graph to demonstrate Case 2: the minimization of the eigenratio $\lambda_n/\lambda_2$. The corresponding schematic for the distributed algorithm can be seen in Figure S3.

The eigenratio decreases rapidly below the threshold for the local stability of the synchronized state for $x$-diffusively coupled Rössler oscillators (dashed line). However, it can be seen that edge weights continue to exhibit small amplitude oscillations at steady state. These oscillations are due to the optimal Laplacian possessing non-distinct extremal eigenvalues.

Specifically, due to the limitation of the distributed estimator for $\lambda_n$ only being able to estimate an associated eigenvector, rather than the eigenspace, weights will adapt to reduce $\lambda_n$ so that it now becomes $\lambda_{n-1}$, and the distributed estimator takes time to converge on the new $\lambda_n$. In this time where the estimator provides an incorrect value, edge weights may overshoot, and set up a persistent oscillation around the optimal value. As the speed of the distributed estimator layers is increased relative to the weight adaptation layers, oscillations become smaller in amplitude and higher in frequency.

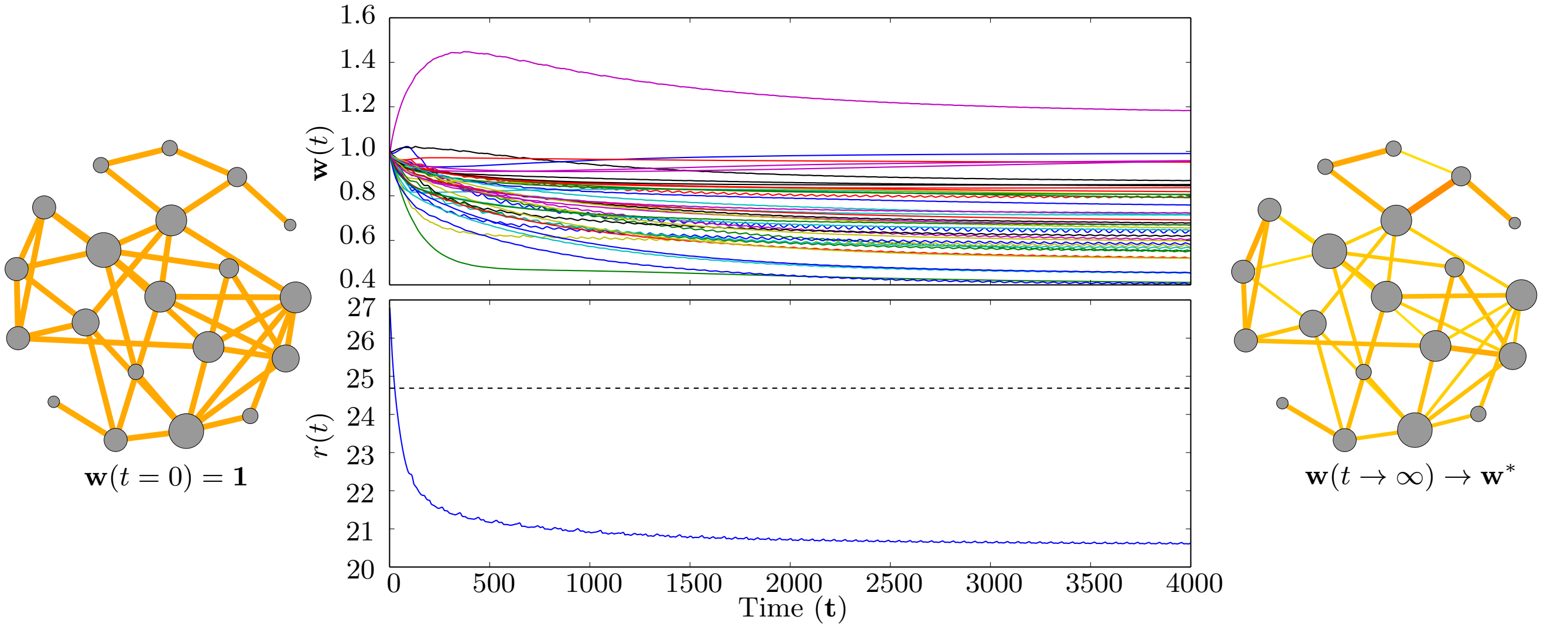

FIG. S 4: Edge weights are adapted in time according to our distributed algorithm for the minimization of $r := \lambda_n/\lambda_2$. It can be seen that the eigenratio $r$ decreases over time settling into a persistent oscillation. Again, in the network diagrams, edge thickness and redness is proportional to edge weight, and node diameter is proportional to the maximum allowed weighted degree at each node.

The presence of these oscillations does not affect in any case convergence towards an optimal value. In practice each edge weight can be "locked" to its average steady-state value as is typically done in the practical implementation of adaptive controllers, e.g.,[5].

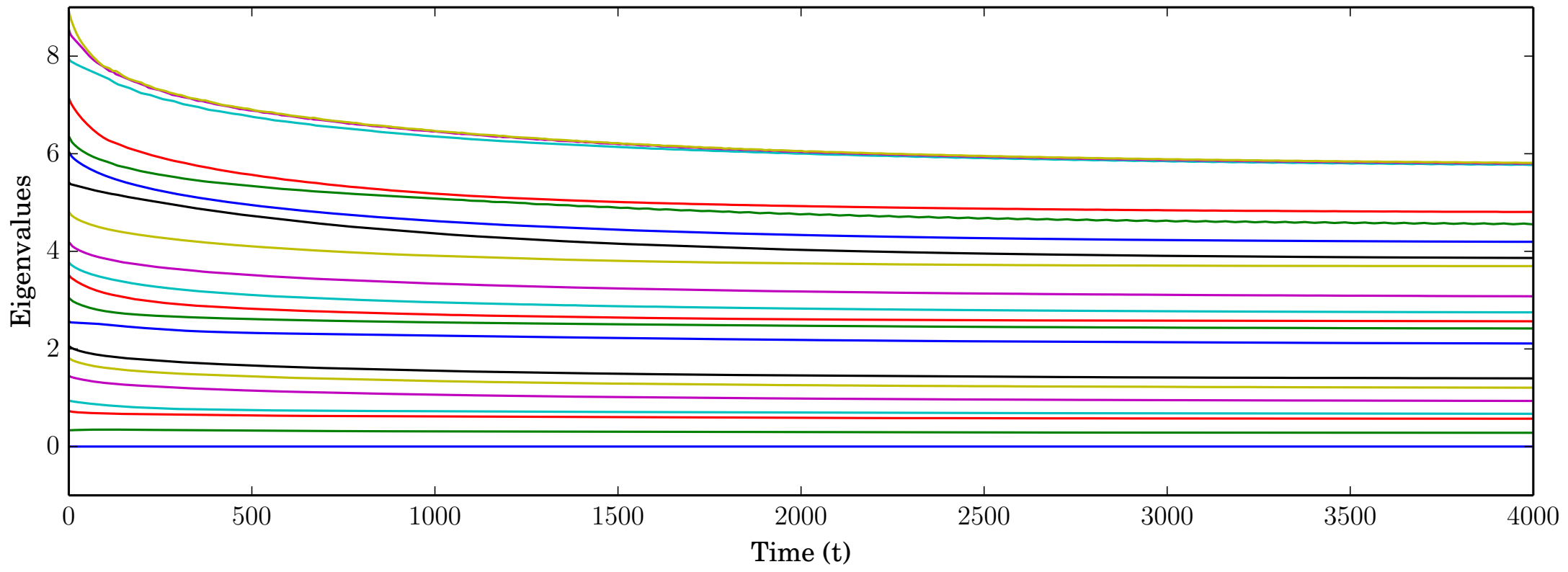

FIG. S 5: Evolution of the eigenvalues of the Laplacian as weights are adapted. It can be seen that as the largest eigenvalue diminishes, it converges on the second and third largest eigenvalues, and in the steady-state, the three largest eigenvalues are non-distinct.

* l.kempton@bristol.ac.uk
† g.herrmann@bris.ac.uk
‡ m.dibernardo@bris.ac.uk